%% file: main.tex
\documentclass{article} 
\usepackage{iclr2023_conference,times}

\input{math_commands.tex}

\usepackage{hyperref}
\usepackage{url}
\usepackage{float}

\title{CommsVAE: Learning the brain's macroscale communication dynamics using coupled sequential VAEs}

\author{Eloy Geenjaar\\
TReNDS center \\
Georgia Institute of Technology\\
Atlanta, GA, USA \\
\texttt{egeenjaar@gatech.edu} \\
\And
Noah Lewis \\
TReNDS Center \\
Georgia Institute of Technology \\
Atlanta, GA\\
\And
Amrit Kashyap \\
Department of Brain Simulation \\
Charite University Hospital \\
Berlin, Germany \\
\And
Robyn Miller \\
TReNDS Center \\
Georgia State University \\
Atlanta, GA \\
\AND
Vince Calhoun \\
TReNDS Center \\
Georgia Institute of Technology \\
Atlanta, GA \\
}

\usepackage{graphicx}
 \iclrfinalcopy 
\begin{document}

\maketitle

\begin{abstract}
Communication within or between complex systems is commonplace in the natural sciences and fields such as graph neural networks. The brain is a perfect example of such a complex system, where communication between brain regions is constantly being orchestrated. To analyze communication, the brain is often split up into anatomical regions that each perform certain computations. These regions must interact and communicate with each other to perform tasks and support higher-level cognition. On a macroscale, these regions communicate through signal propagation along the cortex and along white matter tracts over longer distances. When and what types of signals are communicated over time is an unsolved problem and is often studied using either functional or structural data. In this paper, we propose a non-linear generative approach to communication from functional data. We address three issues with common connectivity approaches by explicitly modeling the directionality of communication, finding communication at each timestep, and encouraging sparsity. To evaluate our model, we simulate temporal data that has sparse communication between nodes embedded in it and show that our model can uncover the expected communication dynamics. Subsequently, we apply our model to temporal neural data from multiple tasks and show that our approach models communication that is more specific to each task. The specificity of our method means it can have an impact on the understanding of psychiatric disorders, which are believed to be related to highly specific communication between brain regions compared to controls. In sum, we propose a general model for dynamic communication learning on graphs, and show its applicability to a subfield of the natural sciences, with potential widespread scientific impact.
\end{abstract}

\section{Introduction}
\label{sec:introduction}
Characterizing macroscale communication between brain regions is a complex and difficult problem, but is necessary to understand the connection between brain activity and behavior. The effect of neural systems on each other, or connectivity, has also been linked to psychiatric disorders, such as schizophrenia \cite{friston2002dysfunctional}. Hence, gaining a deeper understanding of the dynamics underlying the communication between brain regions is both from the perspective of understanding how our brain facilitates higher-order cognition and also to provide insight into and consequently how psychiatric disorders arise. Static functional network connectivity (sFNC) and dynamic functional network connectivity (dFNC), computed respectively as the Pearson correlation between regional activation timeseries over the full scan duration (sFNC) or on shorter sliding windows (dFNC), are among the most widely reported measures of connectivity between brain regions \cite{hutchison2013dynamic}. These approaches calculate the correlation between the timeseries of each brain region to find their coherence, either across the full timeseries (sFNC) or by windowing the timeseries and calculating the correlation within each window (dFNC). Although extensions have been proposed \cite{hutchison2013dynamic}, along with more complex connectivity measures, such as wavelet coherence \cite{yaesoubi2015dynamic} and Granger causality \cite{roebroeck2005mapping, seth2015granger}, Pearson correlation remains the most prevalent measure of brain network connectivity. And even the less commonly employed metrics have issues stemming from some combination of sensitivity to noise, linearity, symmetry, or coarse timescales. Most importantly, these approaches do not directly model the communication, but rather analyze it post-hoc. The pursuit of instantaneous communication between brain regions \cite{sporns2021dynamic} and generative approaches to model communication can potentially lead to models that closely resemble effective connectivity \cite{avena2018communication}. An important advantage of generative models is that they allow us to move away from post-hoc inference from context-naïve metrics toward simulating macroscale brain communication. We propose the use of recurrent neural networks as generative models of communication on both simulated data and functional magnetic resonance imaging (fMRI) data to validate and demonstrate the specificity of our model. Our method complements connectivity metrics, since it does more than quantify the aggregation of communication \cite{avena2018communication}, but directly simulates it, and can thus be analyzed using those same connectivity metrics.

Creating an accurate generative model of the macroscale communication dynamics in the brain is hard, due to its complexity. However, there are some general design principles the brain follows \cite{sterling2015principles}. Generally, the brain tries to minimize its energy use, which is likely also true for communication in the brain. Macroscale communication is an energy-intensive process and involves white matter tracts, which are essentially highways that connect spatially separate parts of the brain. Hence, the amount of information and the number of times information is sent should generally be limited. The bits needed to convey are limited by the brain using sparse coding, which at a lower scale is how neurons encode and communicate information \cite{olshausen2004sparse}. Although mechanistically different, due to metabolic and volume constraints, macroscale communication presumably exhibits strategies similar to sparse coding to efficiently transfer information between neural populations \cite{bullmore2012economy, sterling2015principles}. To incorporate this inductive bias into our model, we regularize the communication from one region in our model to another using a KL-divergence term to a Laplace distribution. The communication itself is modeled as a normal distribution, but by minimizing its divergence with a Laplace distribution we encourage sparser temporal communication. Encouraging sparse temporal communication implies that information is only sent when necessary, and lower rates of communication lead to reductions in the brain's energy requirements.

The goal of our model is to get a better idea of communication dynamics in the brain. To evaluate whether our model is equipped to find the underlying dynamics of a known generative model, we train it on simulated data. After we show that our model finds the correct generative model from the simulated data, we apply our model to neuroimaging data with fairly well-established neural pathways. Although the pathways are well-established, we do not know the exact ground truth communication underlying fMRI data. We do know that communication dynamics depend on the task it is trying to perform, even outside of explicit windows in which these tasks are performed. Hence, to evaluate our model on fMRI data, we train it on three different tasks and use logistic regression to predict what task a small window of timesteps is from using the communication our model predicts. We compare our model to dFNC and show that our method is more sensitive to the task, especially for small windows where no explicit task is performed. This has implications for both resting-state fMRI and task fMRI, since our model may be able to provide a deeper look into how tasks shape macroscale communications in the brain. Understanding task responses allows us to more accurately understand their correlates to behavior and how communication in the brain shapes certain behavior. Since our results indicate that our model is more sensitive to communications even at resting-state that are still related to a task. Being sensitive even outside of external driving of fMRI activity is impactful for resting-state fMRI and psychiatric disorder research, where we expect communication dynamics to be different between controls and patients. Furthermore, since we establish our method on simulated data and then extend it to real data, we provide an initial framework for scientists that may want to extend our method into their area of expertise.

\section{Method}
\label{sec:method}
\paragraph{Generative communication models}
A generative communication model should meet the following requirements: it should not be prone to noise, model the directionality of the communication, and allow for instantaneous communication dynamics. Regarding noise, macroscale communication in the brain is often assessed using fMRI, which is prone to oscillatory sources of noise such as respiration \cite{lund2006non} among others. Connectivity metrics based on correlations are more prone to oscillatory noise since they are global and thus affect the synchrony between fMRI signals in different regions at the same time \cite{birn2012role}. There are generally, however, many reasons why fMRI signals can exhibit similar activation patterns in multiple regions that may or may not always be explained by communication \cite{birn2012role}. In our model, we model the more global signal of each region separately from the communication dynamics. The general trend of the signal is modeled by some initial inputs, and the sparsity regularization on the communication helps push our model towards solutions that model fewer spurious correlations. Secondly, the symmetry of many connectivity metrics has been mentioned previously, and we solve this issue by having each region explicitly send directional communication to each other region. Thirdly, connectivity metrics often estimate the connectivity over a temporal interval. For example, a correlation between two brain regions is calculated over several timesteps $T$. This leads to a trade-off between temporal resolution, and accuracy of the estimated correlation coefficient. In fact, the maximum temporal resolution is two timesteps, since it is impossible to calculate the variance over less than two samples. Our method bypasses this issue by explicitly sending communication at each timestep. Thus, the temporal resolution of the communication we estimate is the same as the temporal resolution of the input signal.

\begin{figure}[h]
    \centering
    \includegraphics[width=\textwidth]{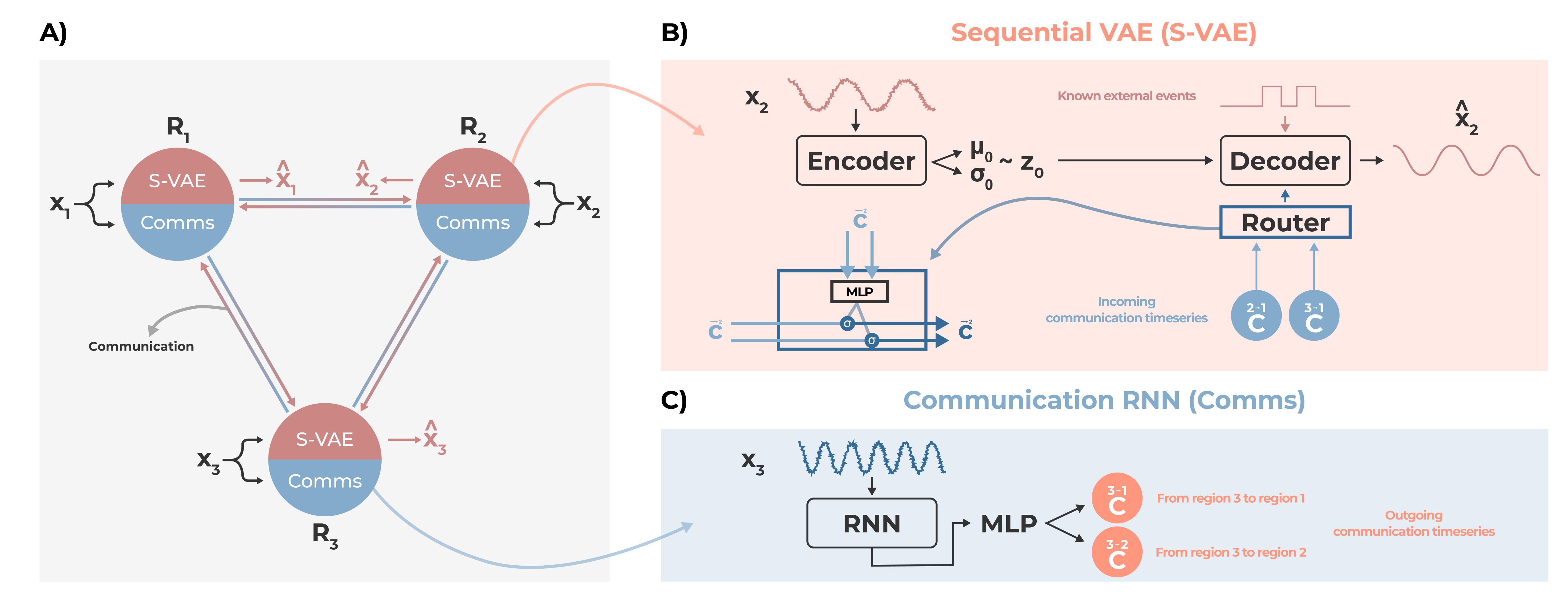}
    \caption{This figure shows the composition of a single node from a graph on the left. The communication and reconstruction RNNs are completely independent of each other, the communication RNN sends signals to other nodes, whereas the reconstruction RNN reconstructs the original timeseries from outside communication and an initial condition $z_{0}$ (right).}
    \label{fig:model}
\end{figure}

\paragraph{Formal definition of our model}
The data structure we assume in this work is a simple graph $\mathcal{G} = (V, E)$, with $N$ vertices ($V$), and $M$ edges ($E$), as shown in Figure \ref{fig:model}A. Generally, we can assume a fully-connected graph, since our proposed model can learn to not send any communication over an edge, meaning it essentially 'trims' that edge. Each vertex $v^{i}$ has a corresponding timeseries $\textbf{x}^{i}_{1...T}$ with $T$ timesteps, this timeseries is modeled as the following dynamical system.
\begin{equation}
    \dot{\textbf{x}}^{i} = \textbf{F}^{i}(\textbf{x}^{i}(t), \textbf{u}^{i}(t))
\end{equation}

One recent model that has been used to model this dynamical system is LFADs \cite{pandarinath2018inferring}, where they model the external inputs $\textbf{u}(t)$ as samples from a normal distribution parameterized by a controller that has access to the hidden state of the encoder and decoder. We want to incorporate a more explicit inductive bias in our model, however. In our case, the external inputs for vertex $v^{i}$ are communications $c^{\to i}_{1...T}$ and external events $\textbf{e}(t)$. We use the notation $\to i$ to mean all incoming communication from neighbors of vertex $v^{i}$, $i \to$ to mean all outgoing communication from vertex $v^{i}$ to its neighbors, and $j \to i$ to mean communication from vertex $v^{j}$ specifically to vertex $v^{i}$. The external events can be known external perturbations of the dynamical system, but are not modeled using a controller like in LFADs \cite{pandarinath2018inferring}. We model the non-linear function $\textbf{F}^{i}$ for each vertex separately, using a recurrent neural network (RNN), with initial inputs $\textbf{z}^{i}_{0}$. We find the initial inputs to the dynamical systems using an encoder, with the following probability distribution 
$\log p(\textbf{z}^{i} | \textbf{x}^{i}_{1...T}) = \log \mathcal{N} \left( \textbf{z}^{i}; \boldsymbol{\mu}^{i}_{0}, \boldsymbol{\sigma}^{i}_{0} \right)$. This distribution is parameterized using a bi-directional RNN, and we sample the initial distributions from it $\textbf{z}_{0} \sim p(\textbf{z}^{i} | \textbf{x}^{i}_{1...T})$. Together with the decoder $p(\hat{x}^{i}_{1...T} | \textbf{z}^{i}_{0}, \textbf{c}^{\to i}_{1...T}, \textbf{e}_{1...T})$, which is parameterized as an RNN as well, the encoder-decoder pair forms a sequential variational autoencoder (S-VAE) \cite{zhao2017towards, kingma2013auto}, except we have added two extra inputs, namely the communication from other vertices, and external events. The S-VAE is visualized in Figure \ref{fig:model}B. Lastly, the communication from one vertex to another is also modeled as a recurrent neural network, with $\textbf{c}^{i}$ as a random variable. $\log p(\textbf{c}^{i} | \textbf{x}^{i}_{1...t}) = \log \mathcal{N} \left(\textbf{c}^{i}; \boldsymbol{\mu}^{i \to j}_{t}, \boldsymbol{\sigma}^{i \to j}_{t} \right)$. Each communication from vertex $i$ to vertex $j$ is thus sampled from this distribution $\textbf{c}^{i \to j} \sim p(\textbf{c}^{i} | \textbf{x}^{i}_{1...t})$. This allows us to regularize the communication distribution between each vertex to be a zero-mean unit-variance multi-dimensional Gaussian, as with variational autoencoders \cite{kingma2013auto}. The communication network is visualized in Figure \ref{fig:model}C. Since each communication network predicts communication to another brain region independently, we also add a router network that works similarly to the forget gate in a long short-term memory (LSTM) network. The controller receives all the incoming communication $\textbf{c}^{\to i}_{t}$ at timestep $t$, these are fed through a multi-layer perceptron (MLP), which has the same number of outputs as the size of the communication vector $\textbf{c}^{\to i}_{t}$. The output is transformed through a sigmoid to be between [0, 1] and multiplied with the incoming communication vector. This allows the router to block information from certain a certain vertex that is already captured by information sent from other vertices. The controller is visually represented in Figure \ref{fig:model}A.

With simplified notation, we thus have the following equations for the S-VAE.
\begin{align}
    \textbf{h}^{i}_{T} &= \text{RNN}^{i}_{\text{Enc}}(\textbf{x}^{i}_{1...T}) \\
    \boldsymbol{\mu}^{i}_{0} &= \textbf{W}^{i}_{\mu}\textbf{h}^{i}_{T}, \qquad
    \boldsymbol{\sigma}^{i}_{0} = \textbf{W}^{i}_{\sigma}\textbf{h}^{i}_{T}, \qquad
    \textbf{z}^{i}_{0} \sim \mathcal{N} \left(\boldsymbol{\mu}^{i}_{0}, \boldsymbol{\sigma}^{i}_{0} \right) \\
    \hat{\textbf{x}}^{i}_{1...T} &=  \text{RNN}^{i}_{\text{Dec}}(\textbf{z}^{i}_{0}, \; \textbf{c}^{\to i}_{1...T}, \; \textbf{e}_{1...T}) \\
\intertext{Then, for the communication network we obtain the following equations.}
    \textbf{h}^i_{t} &= \text{RNN}^{i}_{\text{Comm}}(\textbf{x}^{i}_{\leq t}) \\
    \boldsymbol{\mu}^{i \to}_{t} &= \text{MLP}^{i}_{\text{Comm}, \mu}(\textbf{h}^{i}_{t}), \qquad
    \boldsymbol{\sigma}^{i \to}_{t} = \text{MLP}^{i}_{\text{Comm}, \sigma}(\textbf{h}^{i}_{t}), \qquad
    \textbf{c}^{i \to}_{t} \sim \mathcal{N} \left(\boldsymbol{\mu}^{i \to}_{t}, \boldsymbol{\sigma}^{i \to}_{t} \right) \\
    \textbf{c}^{i \to} &= \textbf{c}^{\to i} \odot \sigma \left(\text{MLP}^{i}_{\text{Router}} \left(\textbf{c}^{\to i} \right) \right)`
\end{align}
Where $\sigma(\cdot)$ is the sigmoid function, and $\text{MLP}^{i}_{\text{Comm}, \mu}$ and $\text{MLP}^{i}_{\text{Comm}, \sigma}$ share every layer, except the last one.intense

We can then adapt the evidence lower-bound (ELBO) \cite{kingma2013auto} as follows, where $q_{\phi}^{i} \left(\textbf{z}^{i}_{0}, \textbf{c}^{i \to}_{1...T} \; | \; \textbf{x}^{i}_{1...T} \right)$ refers to the approximate posterior, and $p_{\theta}^{i} \left(\textbf{x}^{i}_{1...T} \; | \; \textbf{z}^{i}_{0}, \textbf{c}^{\to i}_{1...T} , \textbf{e}_{1...T} \right)$ is the generative network, in our case the recurrent decoder.
\begin{align}
    \mathcal{L}(\theta, \phi, \textbf{x}_{1...T}) &:=
    \sum_{i=1}^{N} \frac{1}{N} \bigg{[} -\text{D}_{\text{KL}}
    \left(q_{\phi}(\textbf{z}^{i}_{0} \; | \; \textbf{x}^{i}_{1...T}) \; || \; p(\textbf{z}) \right) \\
    &- \lambda \text{D}_{\text{KL}} \left(q_{\phi}(\textbf{c}^{i \to}_{1...T} \; | \; \textbf{x}^{i}_{1...T}) \; || \; p(\textbf{c}) \right) \\
    &+ \mathbf{E}_{q_{\phi}^{i} \left(\textbf{z}^{i}_{0}, \textbf{c}^{i \to}_{1...T} | \textbf{x}^{i}_{1...T} \right)} \left[
    \log p_{\theta}^{i} \left(\textbf{x}^{i}_{1...T}\; | \;  \textbf{z}^{i}_{0}, \textbf{c}^{\to i}_{1...T} , \textbf{e}_{1...T} \right) \right] \bigg{]}
\end{align}
Where $p(\textbf{z})$ and $p(\textbf{c})$ are the priors for the initial inputs and the communications and $\lambda$ is the sparsity multiplier, respectively. In our work, we take a zero-mean unit-variance multi-dimensional Gaussian distribution for the initial inputs and a zero-mean $0.1$ standard deviation Laplacian for the communications. This ensures that the model sends communication sparsely to other regions.

\begin{figure}[h]
    \centering
    \includegraphics[width=0.7\textwidth]{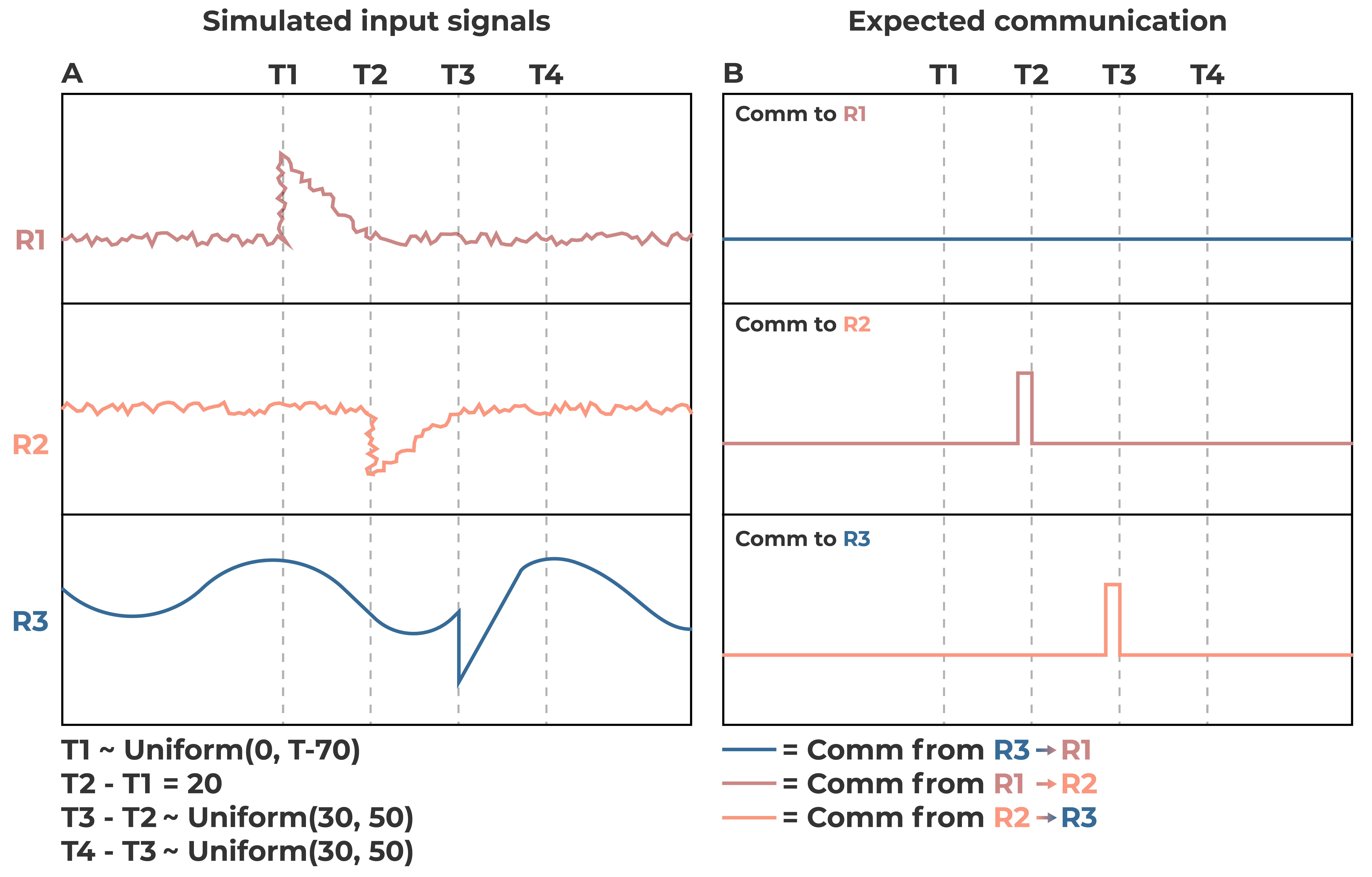}
    \caption{This figure shows the simulated input signals on the left and the communication we would expect our model to find. The communication would happen from R1 to R2 (middle right) right before T2, and from R2 to R3 (bottom right) right before T3.}
    \label{fig:simulated-data}
\end{figure}

\paragraph{Intuition and synthetic data}
To gain some intuition about our model, we explain how our model works on the simulated data we use in this work. Throughout its training process, each region's communication model will try to find temporal patterns in its region that can help reconstruct another region's timeseries. We assume that if this is the case, namely that part of the signal is explained by a pattern in another region at a previous timestep, then communication could have occurred. A simple example is a communication between two regions, where the timeseries for each region is uniform noise. Let us assume there is a linearly decaying pulse in the first region at a time $T1$ ~ $\text{Uniform}(0, 70)$, that decays until $T2 ~ T1 + \text{Uniform}(30, 50)$. Once the pulse has decayed in the first region at time $T2$, a pulse starts in the second region that linearly decays at time $T3 ~ T2 + \text{Uniform}(30, 50)$. These are exactly the parameters we use to create our synthetic data. Now, if we only observe one of the two timeseries, the pulses will seem sampled from a uniform distribution to the decoder of that region. However, if we have access to both regions at the same time and enough training data, our model should be able to figure out that the pulse in the first region structurally precedes the pulse in the second region. In our synthetic data, we add a third region that models a sinusoid and has a pulse after the second region's pulse that linearly decays at time $T4 ~ T3 + \text{Uniform}(30, 50)$. This region also has a different underlying signal, namely a randomly shifted sinusoid (shifted between $~ \text{Uniform}(-20, 20)$) to show that our model can separate global signal from communication. A visual example of our synthetic data is shown in Figure \ref{fig:simulated-data}A. We expect that the initial inputs $z_{0}$ models the global signal so that the communication our model predicts is not affected by the sinusoidal signal. As mentioned previously, fMRI data also exhibits global signal that may affect connectivity metrics, such as physiological noise. Hence, we can infer whether our model is affected by the global signal in our simulation data to make it more realistic when transferred to biological data. The communication we thus expect is shown in Figure \ref{fig:simulated-data}B. Namely, the communication model in region 1 should learn to send a signal to region 2 when the pulse in its region has almost decayed, and region 2's communication model should learn to send a pulse to region 3 once its pulse has almost decayed. The expected communications in our synthetic data are shown in Figure \ref{fig:simulated-data}B. To evaluate whether our model finds the expected communication, we center a normal distribution at the start of the pulse for regions $2$ and $3$. We then correlate the communication from region $1$ to region $2$ and region $2$ to region $3$ with that normal distribution for multiple shifts between $-20$ and $2$. We take the maximum correlation and subtract two times the L1 sparsity of the other communications from it to obtain a measure of similarity, where higher is better. 

\paragraph{Biological data} 
\label{sec:biological-data}
The fMRI data we use in this work is obtained from the open-access, under data usage terms, HCP-1200 dataset \cite{van2013wu} for all subjects with cortical surface timeseries data. We use three tasks in our work, the Motor, Emotion Processing, and Language Processing tasks. The data is registered using multimodal surface registration (MSM) \cite{robinson2014msm, robinson2018multimodal}, and surfaces are constructed using Freesurfer \cite{glasser2013minimal, fischl2012freesurfer}. We use the Yeo-17 atlas \cite{yeo2011organization} to separate the cortical surface data into $17$ separate regions. The spatial information within each region is averaged, as is commonly done in communication analyses of the brain. Each subject's timeseries is then independently band-pass filtered independently ($0.01-0.15$Hz) and linearly detrended using Nilearn package \cite{abraham2014machine}. Each task has a different length, so we cut off tasks at $176$ timesteps, which is the smallest common size of the timeseries. The motor task consists of a visual cue, after which a subject has to start performing a motor task. The motor tasks a subject performs are related to the left hand, right hand, left foot, right foot, and tongue. For the emotion task, the subject sees a visual cue and then has to match two shapes shown at the bottom of the screen to a shape at the top. This sub-task is referred to as the neutral task, and the other task has the subject match one of two faces at the bottom of the screen with a face shown at the top. The faces either look fearful or angry. The language task consists of a sub-task where the participant has to perform arithmetic and push a button for the right answer. The story sub-task presents the subjects with a short story of 5-9 sentences. The timing of each task is shown, after convolving it with SPM's hemodynamic response function \cite{penny2011statistical}, in Appendix \ref{app:hemodynamic}. We train one model on all three tasks such that the prediction of the task is not influenced by the fact that the model was specifically trained to model communication for that task.

\paragraph{Implementation}
Our model is implemented in Pytorch \cite{paszke2017automatic} and trained on an internal cluster using single NVIDIA V100 and NVIDIA 2080 GPUs, with a batch size of $8$, the Adam optimizer\cite{kingma2014adam}, a 1E-4 weight decay, a learning rate of 0.01, $0.1$ epsilon, and $0.9$, $0.999$ as betas. We reduce the learning rate when it plateaus using a scheduler. At each plateau, we reduce the learning rate by multiplying it with $0.95$, with a patience of $6$ epochs, and a minimum learning rate of $1E-5$. We train each model for $200$ epochs, for the simulated data across $8$ seeds: ($42$, $1337$, $9999$, $50$, $100$, $500$, $1000$, $5000$). The model is trained on the first three seeds for the biological data, this is due to computational limitations. Since we use multiple sparsities, the number of runs quickly explodes, and each model on biological data takes about $5-6$ hours to train. The hidden size we use for the encoder, decoder, and communication model are all $16$, and the latent dimensions for $z_{0}$ is $32$. For the synthetic data we use sparsities: ($0.001$, $0.025$, $0.005$, $0.0075$, $0.01$, $0.025$, $0.05$, $0.075$, $0.1$, $0.25$, $0.5$, $0.75$, $1.0$, $2.5$). For the real data we use a smaller number of sparsities due to computation limitations, but also because we have narrowed the sparsity down at this point to: ($0.001$, $0.005$, $0.01$, $0.025$, $0.05$, $0.1$, $1.0$, $2.0$, $3.0$). In order to not mix subjects across the training, validation, and test set, we split them based on subject. Hence, a subject occurs with every task in one of the sets it was assigned to and in no other sets.

\paragraph{Experiments}
To test whether our model is interesting as a complement to connectivity metrics, we evaluate our model in three different ways. To get a better idea of how much the controller improves our model, we perform each experiment with both models. We refer to the model without the controller as CommsVAE and the model with the controller as Meta-CommsVAE in the rest of the text. Firstly,  we evaluate our model on the aforementioned synthetic data. We calculate the correspondence between the expected communications and the communication that our model predicts on a test set and look at how the sparsity parameter affects the average correspondence to the test set. Note that when we use the communication in these experiments, we take the mean of the Gaussian distribution, since it is the most probably point. Then, we train both our models on an fMRI dataset with three tasks, each model is trained on all three tasks conjointly not to induce any task-specific model noise that can affect the downstream prediction. Since we do not have a ground truth for the communication of the fMRI signal, we determine the best model by looking at the reconstruction error on the validation set with respect to the sparsity, see Appendix \ref{app:reconstruction-validation}. We use the model that performs the best on average over all the seeds for a certain sparsity parameter and uses it for task prediction. We use $4$ different window sizes [$5$, $10$, $20$, and $50$] to compare our model to dFNC. This allows us to determine whether our model has better temporal resolution since the quality of the correlation will get lower with a smaller dNFC window. For each window, we take the absolute value of our model's communication (since it can be both negative and positive) and sum it across each window. Then, we train a logistic regression classifier with L1 sparsity on the training and validation set using the temporally summed communication to predict what task each subject in the test set's window is from. This results in a classification accuracy across time for the four windows, and we perform the same routine for the dFNC. Finally, we look at the absolute sum of the communication for each task. We average the absolute temporal sum over the subjects in the test set and visualize a graph of the communication that exhibit more than $25\%$ of the maximum sum for each task. The reason we need to take the absolute sum is that the model can send both positive and negative communication.

\section{Results}
\begin{figure}[h]
    \centering
    \includegraphics[width=0.75\textwidth]{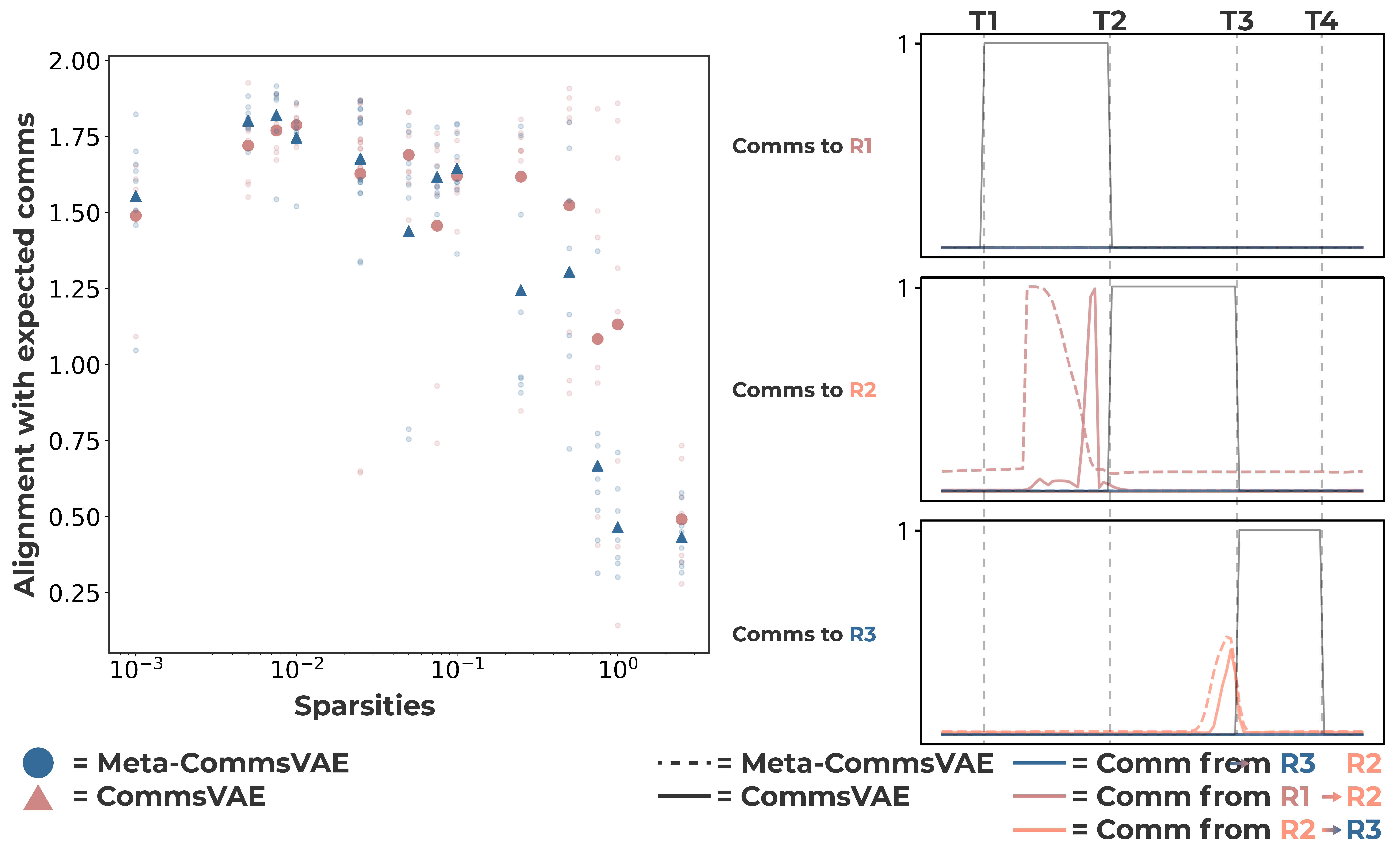}
    \caption{This figure shows the performance of our models for different sparsity parameters (x-axis) on the left and a comparison of our model's communications to the ground truth event times (black) on the right. The values at which events happen are Uniformly sampled, but we shift each timeseries in the test set to align the communication and events and then average the communication over the test set.}
    \label{fig:results-simulated}
\end{figure}

\paragraph{Synthetic data}
The results for the synthetic data are shown in Figure \ref{fig:results-simulated}. Generally, we see a trend where higher sparsity parameters increase the correspondence to our expected communication on the test set, up to a point where increased sparsity suppresses any communication. There is a peak around $0.01$ in Figure \ref{fig:results-simulated}A for both models, after which the correspondence with the expected communication decreases. This exactly supports our hypothesis that encouraging sparse communication helps. To show what the communication looks like in the two models we test, we take the test set synthetic data, predict the communication for each, and use the synthetic generation parameters to align the communication timeseries. After aligning the communication timeseries, we take the average over the test set and visualize what the communication looks like in Figure \ref{fig:results-simulated}B. Right before the pulse ends in the region $1$, it communicates with region $2$ before its pulse begins. The communication from region $2$ to region $3$ starts exactly when the pulse is almost linearly decayed, right before the pulse starts in region $3$. The main difference between our Meta-CommsVAE and CommsVAE model is that in Figure \ref{fig:results-simulated}B, the CommsVAE's communications between region $1$ and region $2$ are closer to the actual pulse. Generally, we do not see a trend in terms of a difference between the two CommsVAE models, at least for the synthetic data. This is probably because the task itself is sparse and hence being able to increase sparsity using the controller is unnecessary. An example of the reconstruction of validation data by our model is shown in Appendix \ref{app:reconstruction}.

\begin{figure}[h]
    \centering
    \includegraphics[width=0.8\textwidth]{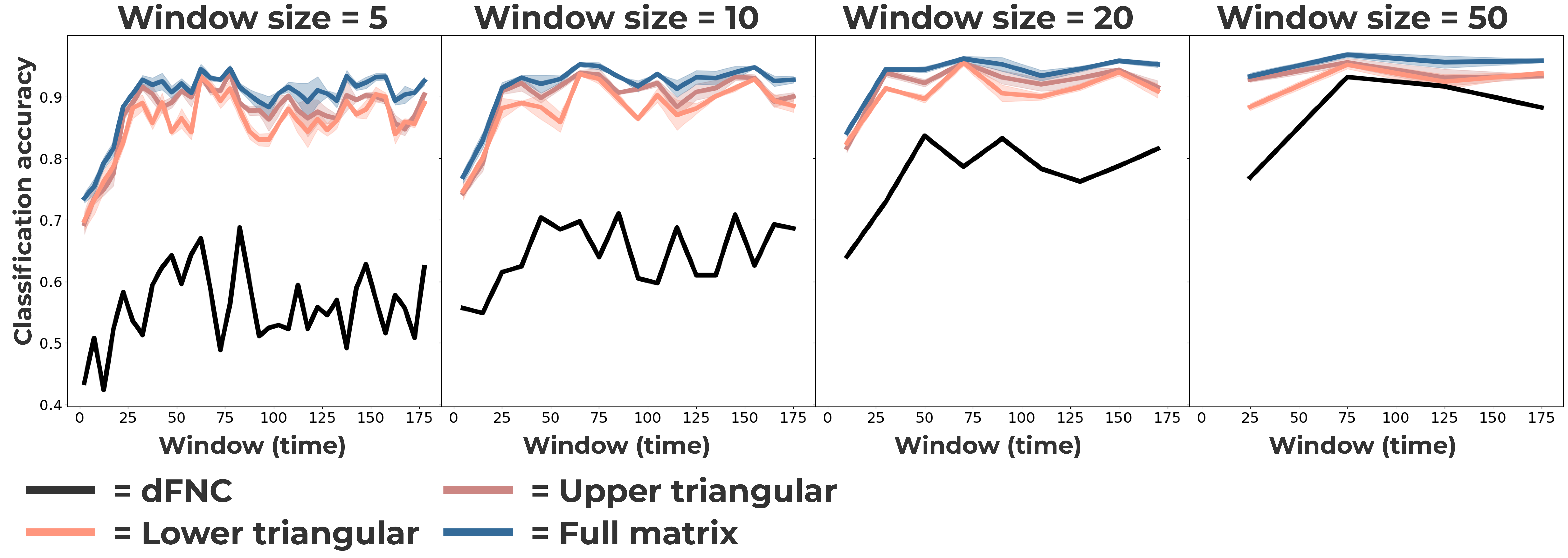}
    \caption{This figure shows four panels, each with a different window size to compute both our model's communication density and dFNC. The y-axis shows the classification accuracy of the task given only that window on the test set.}
    \label{fig:results-fmri}
\end{figure}

\paragraph{fMRI data}
\label{sec:fmri-data}
The results to test both the specificity of our generative model and how instantaneous the communication is are shown in Figure \ref{fig:results-fmri}. We produce these results for the best model we find based on the reconstruction error on the validation set, see Appendix \ref{app:reconstruction-validation}. The best model is the Meta-CommsVAE model with a sparsity of $0.025$. For the next two results, we will use this model. Our results with this model clearly show that our method outperforms dynamic functional connectivity (dFNC) in terms of predicting what task each window is from. This is true for every window, and every window size. Furthermore, our method only has a small decline in classification accuracy for smaller window sizes, which supports our hypothesis that our model can find instantaneous communication. Our method already starts being task-specific early on in time, meaning that even before any of the tasks start, the communication our model finds in the brain is task-specific, which is likely to be true if we knew the ground truth of communication. One reason that the model increases in task accuracy is that the communication model is an RNN and only has access to points previously, so it does not have enough information to start sending information earlier on in the timeseries. After a few windows, the accuracy of the task increases and is then relatively stable. This is different for the dFNC, which fluctuates, likely based on when the tasks start, but our model is specific to the task essentially the same across all of the windows in the timeseries. Thus, our model is even task-specific when there is relatively little task signal from either of the tasks in a certain window. Another result we can gather from Figure \ref{fig:results-fmri} is that the lower-triangular part of the matrix is worse at predicting the task than the upper triangular part of the communication matrix for each window. The reason we include these results is that the dFNC is symmetric and the lower and upper triangular parts of the matrix are thus the same. To make the comparison fair, we compare separately compare the triangular parts of our communication matrix with dFNC. Since our method is directed, we can also take the whole matrix, which outperforms both the lower and upper triangular parts of that matrix, as expected.

\begin{figure}[h]
    \centering
    \includegraphics[width=0.9\textwidth]{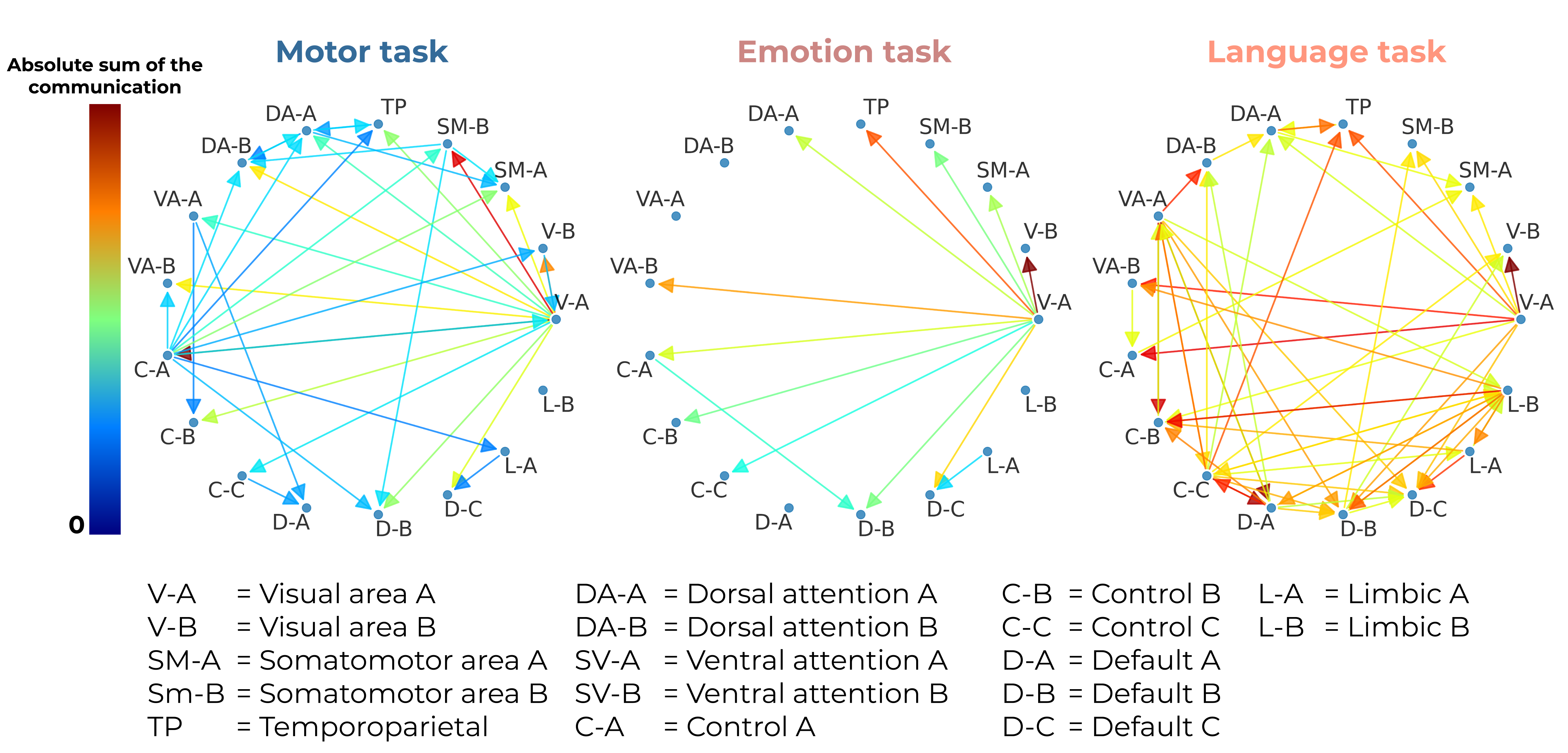}
    \caption{This figure shows the absolute temporal sum over the predicted communication dynamics from our model. The colors are scaled within each task, and the colorbar goes from $0$ to the maximum of the absolute sum for each task. The reason we do not show any numbers is because they are dimensionless.}
    \label{fig:results-interpretation}
\end{figure}

Our last experiment aims to evaluate whether our model finds explicitly directed communication that is different across the tasks. The hemodynamic response model over time for each sub-task is shown in Appendix \ref{app:hemodynamic}. The absolute sum over the timeseries for each task is shown in Figure \ref{fig:results-interpretation}. There are a few interesting results that align with known scientific processes. For example, in all models, the communication from the earlier visual areas (V-A) is stronger towards the later visual areas (V-B). Furthermore, the communication from early visual areas (V-A) to the somatomotor regions (SM-B) is the strongest in the motor task. Since the motor sub-tasks are preceded by a visual cue, it is expected that the early visual areas send information to the somatomotor areas, since those have to execute the task. Another interesting result is that one of the control regions (C-A) in the motor task is an important node in sending information to other brain regions. Generally, most regions only send to other regions, without receiving any communications from those regions, meaning that our model finds explicitly directed communication. This aligns with our hypothesis that generative models should be directed. This is especially obvious for the emotion task, where communication from the early visual areas (V-A) to other regions is the most dominant. The emotion task also contains a visual cue. The visual communication in the language task is less strong compared to the other regions, and there are many other communication dynamics, but since the math sub-task is partly visual, we do expect some of the visual communication that we observe in the figure. One of the large differences between the motor and emotion, and language task is that the limbic system is much more involved in the language task. We would potentially expect some limbic communication in the emotion task as well since one of the sub-tasks pertains to seeing angry/fearful faces. The limbic system composes the hippocampus and amygdala, which are involved in the storage of memories and regulating emotional memories, respectively. The reason that the language task could evoke communication from the limbic system is thus that the stories evoke some emotional memories in the subjects. The stories are from Aesop's fables and may have modulated some emotional memories in the subjects. However, inferences like these need to be studied more robustly in future work.

\section{Discussion}
This work proposes a generative model for macroscale communication in the brain that we show predicts directed, instantaneous, and task-specific communication dynamics. We compare our model on a synthetic task that tests the requirements we set out for our model and find that our model can accurately uncover the communication dynamics we incorporate into the synthetic data. Then, we train our two models on three fMRI tasks and find that reconstruction error decreases with higher sparsity parameters, indicating that the sparse prior on brain communication dynamics fits our data. The sparsity of macroscale brain communication needs to be studied further but aligns with findings in computational neuroscience \cite{olshausen2004sparse} and the design principles of the brain \cite{sterling2015principles}. We use a model with the lowest reconstruction error to predict what task small time windows are from in a test set and compare the classification accuracy with a commonly used metric, dynamic functional connectivity (dFNC). Our model performs much better with smaller window sizes, supporting our claim that our model predicts instantaneous communication. Then, to understand the added benefit of directed models, we look at the absolute sum of the communication for each task and find that almost all, especially strong, connections are directed. The directions and the strength of the communication also align with neuroscientific priors we have about how regions affect each other. Thus, our results confirm what we set out to incorporate into our model from the beginning.

We want to expand on this model by applying it to resting-state fMRI data, and explore whether our model can be used to more deeply understand psychiatric disorders. Furthermore, since this model is generative, we can start to probe the model and interpret the response functions when we synthetically increase communication from one region to another at certain time points. We can design communication patterns to test how communication in different regions affects the signal in others.

\subsubsection*{Acknowledgments}
Data were provided by the Human Connectome Project, WU-Minn Consortium (Principal Investigators: David Van Essen and Kamil Ugurbil; 1U54MH091657) funded by the 16 NIH Institutes and Centers that support the NIH Blueprint for Neuroscience Research; and by the McDonnell Center for Systems Neuroscience at Washington University. This material is supported by the National Science Foundation under Grant No. 2112455 and the National Institutes of Health grant \#R01MH123610. Eloy Geenjaar was supported by the Georgia Tech/Emory NIH/NIBIB Training Program in Computational Neural-engineering (T32EB025816).

\bibliographystyle{iclr2023_conference}
\bibliography{iclr2023_conference}

\appendix
\newpage
\section{fMRI data reconstruction error across sparsity}
\label{app:reconstruction-validation}
\begin{figure}[h]
    \centering
    \includegraphics[width=\textwidth]{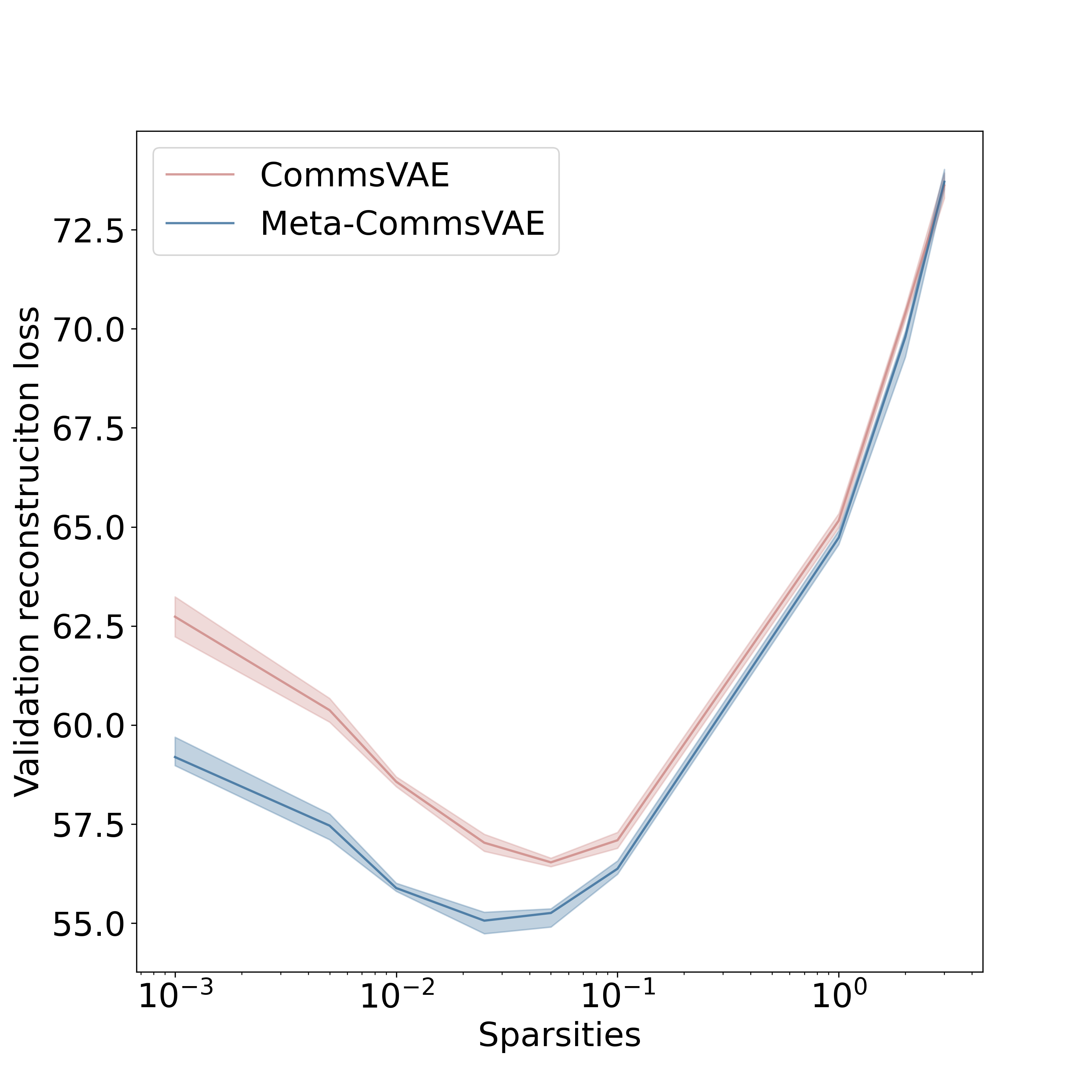}
    \caption{This figure shows the average reconstruction of both models on the fMRI dataset, with standard deviation computed across the seeds. The Meta-CommsVAE clearly outperforms the CommsVAE for all sparsities, and performs the best at a sparsity of $0.025$.}
    \label{fig:validation-reconstruction}
\end{figure}

Figure \ref{fig:validation-reconstruction} shows that the Meta-CommsVAE outperforms the CommsVAE on all sparsities, but also has a clear dip around a certain sparsity. Since we do not know the ground truth of the communication in the fMRI dataset like we did with the synthetic data, the improvement in reconstruction error for higher sparsities shows promise for hyperparameter searches on other datasets. Note the reconstruction error is a squared error, summed over time and averaged over the number of regions.

\newpage
\section{Tasks hemodynamic response}
\label{app:hemodynamic}
\begin{figure}[h]
    \centering
    \includegraphics[width=\textwidth]{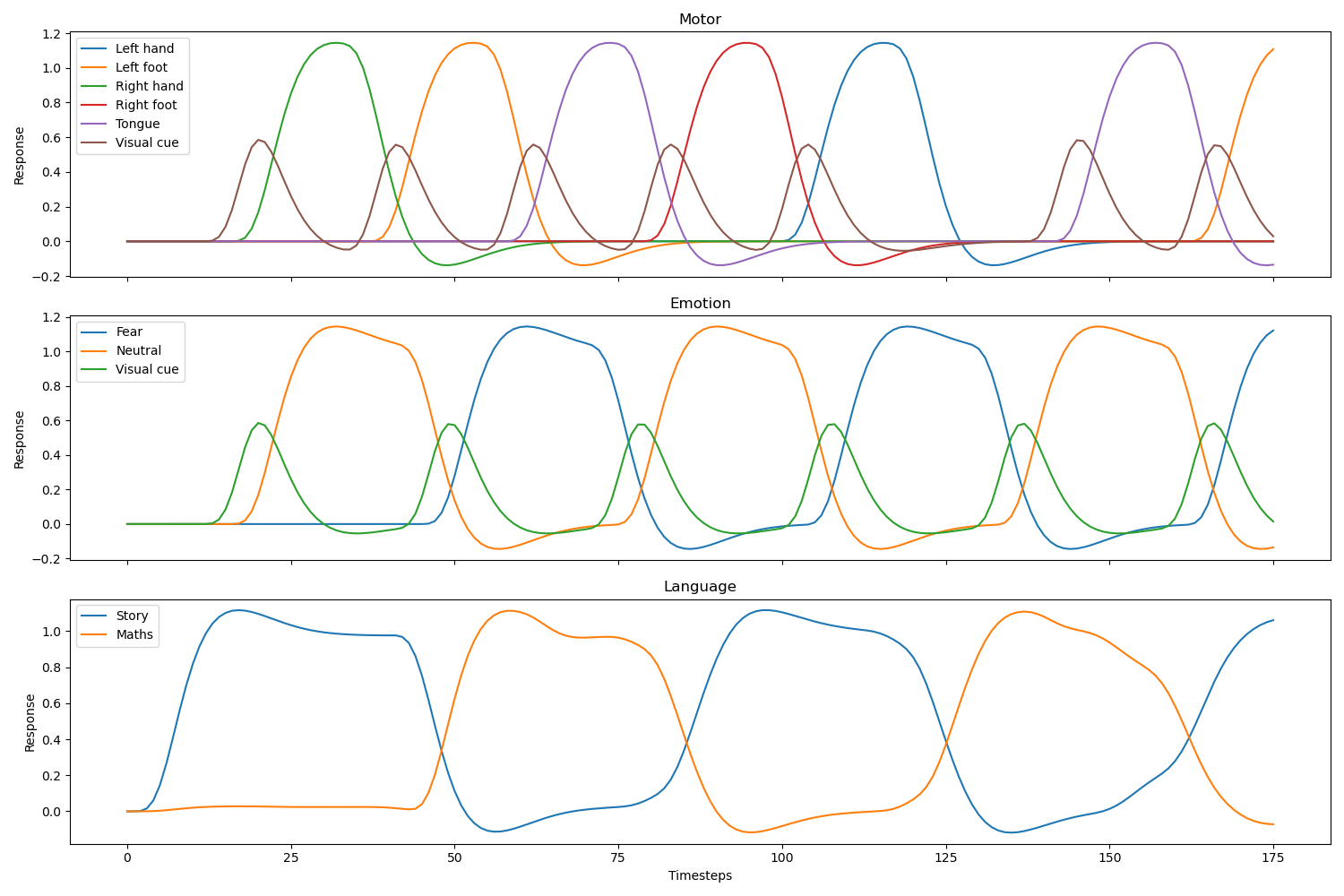}
    \caption{This figure shows the average hemodynamic response of the test set.}
    \label{fig:hemodynamic-response}
\end{figure}

Figure \ref{fig:hemodynamic-response} shows the hemodynamic response for each of the tasks for the $176$ timesteps we use in our model. The legend indicates what sub-task the hemodynamic response corresponds to.

\newpage
\section{Example reconstruction of the synthetic data}
\label{app:reconstruction}
\begin{figure}[h]
    \centering
    \includegraphics[width=\textwidth]{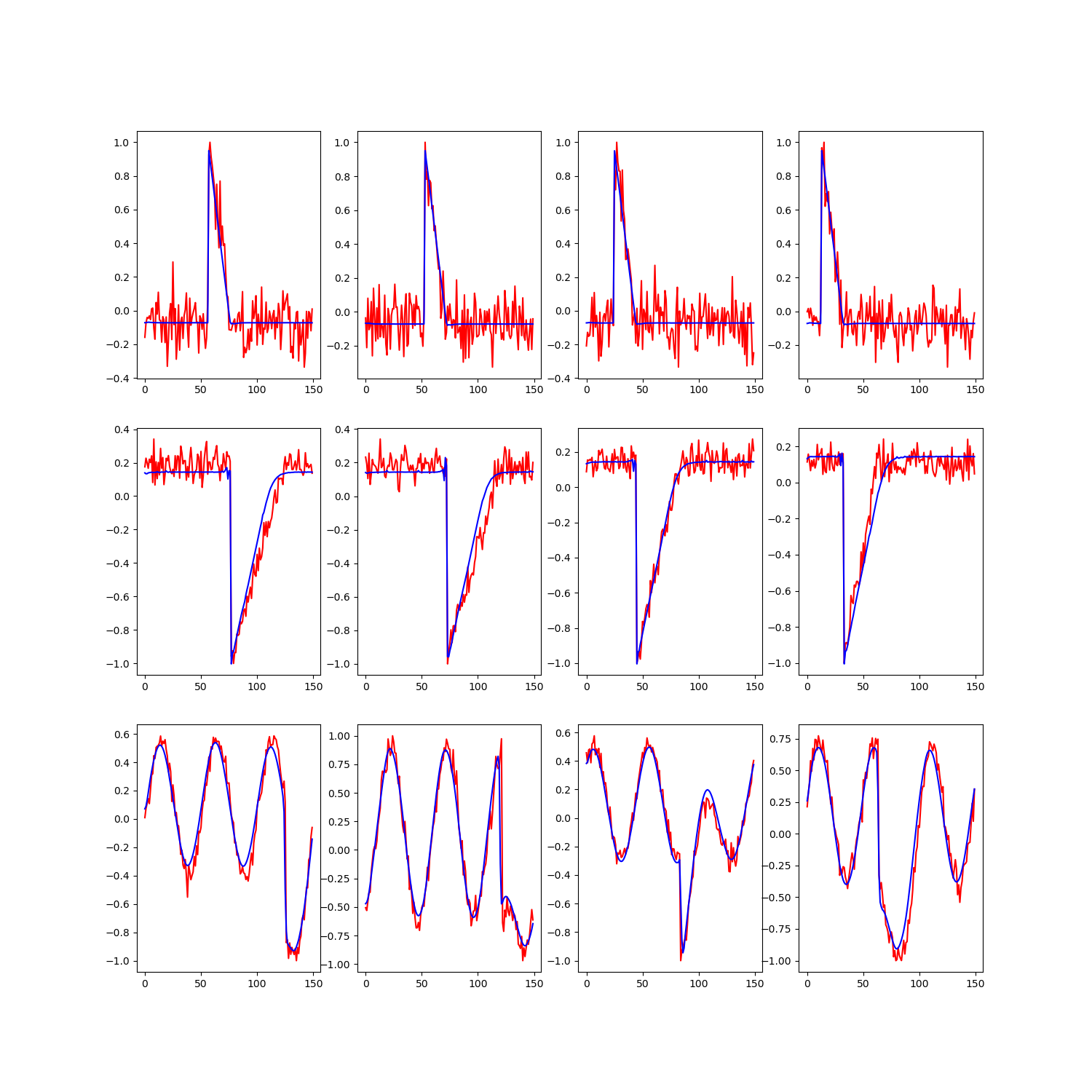}
    \caption{This figure shows an example of the reconstruction on the synthetic data for our CommsVAE model with sparsity $0.01$. The red line is the input signal, the blue line the reconstructed signal, each row corresponds to a different region, and each column to a different timeseries from the validation set.}
    \label{fig:simulated-reconstruction}
\end{figure}

Figure \ref{fig:simulated-reconstruction} shows an example of the reconstruction of the synthetic data. The red line is the input signal, and the blue line is the reconstruction. The first row is the first region, the second row the second region, and the last row the third region. The columns are four random timeseries from the validation set. This shows that our model can still model the global signal, even for a shifted sinusoid, with almost perfect communication dynamics, see Figure \ref{fig:results-simulated},


\end{document}

%% file: math_commands.tex
\usepackage{amsmath,amsfonts,bm}

\def\eqref#1{equation~\ref{#1}}

\def\1{\bm{1}}

\DeclareMathAlphabet{\mathsfit}{\encodingdefault}{\sfdefault}{m}{sl}
\SetMathAlphabet{\mathsfit}{bold}{\encodingdefault}{\sfdefault}{bx}{n}

